\documentclass[aps,prl,twocolumn,groupedaddress]{revtex4}

\usepackage{graphicx}
\usepackage{dcolumn}
\usepackage{bm}

\def\lsim{\raise 0.4ex\hbox{$<$}\kern -0.8em\lower 0.62
ex\hbox{$\sim$}}

\def\gsim{\raise 0.4ex\hbox{$>$}\kern -0.7em\lower 0.62
ex\hbox{$\sim$}}

\def\lbar{{\hbox{$\lambda$}\kern -0.7em\raise 0.6ex
\hbox{$-$}}}

\newcommand\eq[1]{eq.~(\ref{#1})}

\newcommand\ee{\end{equation}}
\newcommand\be{\begin{equation}}
\def\bea{\begin{eqnarray}}
\def\eea{\end{eqnarray}}
\newcommand\ees{\end{eqnarray}}
\newcommand\bees{\begin{eqnarray}}

\def\p1{{\bf p}_1}
\def\p2{{\bf p}_2}
\def\k1{{\bf k}_1}
\def\k2{{\bf k}_2}

\newcommand{\fmax}{f_{\rm max}}
\newcommand{\DE}{E_{\rm gw}}
\newcommand{\msun}{M_{\odot}}

\def\p{\partial}

\def\g{\gamma}

\def\dslash{\hspace{-1mm}\not{\hbox{\kern-2pt $\partial$}}}
\def\Dslash{\not{\hbox{\kern-4pt $D$}}}
\def\pslash{\not{\hbox{\kern-2.1pt $p$}}}
\def\kslash{\not{\hbox{\kern-2.3pt $k$}}}
\def\qslash{\not{\hbox{\kern-2.3pt $q$}}}

\begin{document}


\title{Experimental signatures of gravitational wave bursters}

\author{Florian~Dubath}
\author{Stefano~Foffa}
\author{Maria~Alice~Gasparini}
\author{Michele~Maggiore}
\author{Riccardo~Sturani}

\affiliation{D\'epartement de Physique Th\'eorique, 
Universit\'e de Gen\`eve, 24 quai Ansermet, CH-1211 Gen\`eve 4}

\vspace{5mm}

\date{\today}

\begin{abstract}
Gravitational wave bursters are sources which emit repeatedly bursts
of gravitational waves, and have been recently suggested as potentially
interesting candidates for gravitational wave (GW) detectors.
Mechanisms that could give rise to a GW burster can be found 
for instance in highly
magnetized neutron stars (the ``magnetars'' which explain the
phenomenon of soft gamma repeaters), in accreting neutron stars and in
hybrid stars with a quark core.
We point out that these sources  have very
distinctive experimental signatures. In particular, 
as already observed in the $\g$-ray bursts from soft gamma repeaters, 
the energy spectrum of the events is a power-law,
$dN\sim E^{-\g}dE$ with $\g\simeq 1.6$, and
they have a 
distribution of waiting times (the times between one outburst and the
next) significantly 
different from the
distribution of uncorrelated events.
We discuss possible detection strategies that could be used to
search for these events in existing gravitational wave detectors.

\end{abstract}


\maketitle

{\em Introduction}. 
Astrophysical sources of GWs are traditionally divided into 
continuous  and burst sources.  Continuous sources include, e.g.
spinning neutron stars (NS) or inspiraling binaries, and have a very
long duration.
GW bursts, instead, are due to cataclysmic events and 
can have  a typical duration of
the order of few milliseconds, as for instance 
in supernova explosions or in the
merging phase of a binary NS-NS system. 
In GW detectors, these two classes of  events are therefore
searched using very different  data analysis strategies.

It is normally assumed that a  detectable GW
burst must result from a catastrophic  event that
disrupted the original system. In fact, with the sensitivity of the GW
detectors that are presently taking data~\cite{ROG01,LIGO}, a
GW burst is detectable if it originates from a process that, 
in order of magnitude, liberated in GWs an energy 
\be\label{1Erad}
\DE \sim (10^{-3}-10^{-2}) \, \msun c^2\, \,
(r/8\, {\rm kpc})^2
\, ,
\ee
assuming  a  Fourier spectrum extending up to a maximum frequency
$\fmax\sim 1$~kHz and denoting by $r$ the distance to the source.
As a reference value for $r$ we have taken the distance $r\simeq
8$~kpc 
to the galactic center.
The emission of  $10^{-2}\msun$ in a GW burst is  expected
in a supernova explosion or in the coalescence of a NS-NS
binary so it 
is clear that such a huge energy release 
can only be obtained in a catastrophic event that destroys the
original system.

Smaller GW bursts, instead,  could leave the source in the condition
of flaring again. In particular, 
in a recent paper \cite{CDM} it has been
pointed out that in compact stars there are various
mechanisms that can produce  repeated  activity, with GW
bursts  at the level 
$\DE \simeq 10^{-10}-10^{-8} \msun c^2$, and these mechanisms could 
in fact be
quite common.
These  ``small'' bursts could  turn out to be very
interesting for detection for two reasons. 
First, for sources at typical galactic distances
they could be detectable by
advanced interferometers
like LIGO~II.
Second,   catastrophic events
like supernova explosions are very rare on the galactic scale,
and therefore to have an interesting rate
we should be able to reach out to distances of the order
of the Virgo cluster. Instead,
the smaller burst activity could  be
a rather generic phenomenon in compact stars.
The closest observed
neutron stars are at less than  100~pc. Estimates from
population
synthesis  suggest that the  NS closest to us is  at  a distance 
$r\sim 5-10$~pc and that within a radius of about 100 pc there
should be $O(10^4)$ neutron stars~\cite{ST}. Bursts 
liberating $10^{-8}\msun$, from
sources at such close distances, could 
be interesting  even for present detectors.

As we will discuss in this paper,
the fact that GW bursts could  come repeatedly from the same source
opens new perspectives for the detection strategies, since it allows us
to search for very distinctive correlations between different events, 
both in energy and in the arrival
times.

{\em The physics of GW bursters}.
The mechanisms that can produce a GW burster are generically
associated to a competition between an agent which induces stresses
in a compact star, and some form of rigidity  of the star
itself. When the  stresses mount over a critical value, the
star yields to these  forces and readjusts itself through a
starquake. This releases temporarily the
stresses, which then mount again, resulting in further starquakes.

A very interesting example of this phenomenon is provided by magnetars,
which are neutron stars with huge magnetic
fields, of order $10^{14}-10^{15}$G \cite{DT,TD1}, i.e. 100 to 1000
times stronger than in ordinary pulsars. It is believed that
magnetars provide an explanation for the phenomenon of soft gamma
repeaters (SGR),  x-ray sources with a persistent
luminosity of order $10^{35}-10^{36}$ erg/s, that occasionally emit
huge bursts of soft $\g$-rays, with a power up to $10^{42}$ erg/s, for
a duration of order $0.1$~s.
The mechanism invoked to explain the burst activity is that
the magnetic field lines
in magnetars drift through the liquid interior of the NS, stressing
the crust from below and generating strong shear strains. For magnetic fields
stronger than about $10^{14}$~G, these stresses are so large that
they cause the breaking of the  1~km thick NS crust, whose
elastic energy is suddenly released  in a large
starquake, which generates a burst of soft gamma rays.
As computed in Refs.~\cite{FP2,CDM}, the starquake can radiate
in GWs an energy $\DE\sim 10^{-10}-10^{-9}\msun c^2$. 
Occasionally SGRs emit truly giant flares, and in this case the
estimate for the energy radiated  can be  $ 10^{-8}\msun
c^2$ or even larger~\cite{Ioka}.

Magnetars are just one example of
GW bursters, and a number of variants
have been considered~\cite{CDM}. For instance the trigger, rather than being
provided by the magnetic field, can be provided by accretion onto a NS:
when a critical amount of material 
has been accreted  from a companion, the rigid crust breaks and the star
rearranges itself to a new equilibrium configuration through a
starquake. This liberates 
$10^{-10}-10^{-9}\msun c^2$ of energy, because this is
the maximum elastic energy that can be stored in the crust before
breaking~\cite{Rud,SW} and this energy can be converted in GWs with high
efficency. For a NS accreting mass steadily from a companion, the
process starts  over again, until  a critical mass has  again been accreted
and a new starquake takes place.

Another possibility considered in Ref.~\cite{CDM} is 
a phase transition in the core of a hybrid quark-hadron star. In
this case, each time
the  pressure
in the star interior rises above a critical value, 
for instance because of accretion of matter,
a phase transition takes place, and transforms successive layers of the
star core from the hadronic to the deconfined phase. 
A variant of this scheme uses as a
trigger of the quark-hadron phase transition the spin-down of the NS, which
causes a decreases of the centrifugal force~\cite{DPB}. 
While the rigidity of the crust is determined by the Coulomb
interaction between nuclei, in the core the relevant energy scale is
determined by
hadronic physics, so
corequakes could be a much more powerful source of GWs than  crustquakes
(see also Ref.~\cite{MVP}).

Independently of the specific mechanisms, 
all these models share the features  of
self-organized criticality, which is characterized by the fact that an
agent drives the system toward a critical state, until the energy of
the system is suddenly, and often catastrophically, released. Typical 
members of this class are avalanches, earthquakes, and solar flares
(for a broad overview of self-organized criticality, see Ref.~\cite{Bak}).
Phenomena showing self-organized criticality have a certain degree of
universality, in the sense that their statistical properties are
largely independent of the details of the  physical mechanisms
involved. 
For this reason, it is possible to make some predictions which are very
general, even without a knowledge of the  details
of the system. In the following we concentrate on two
aspects which could be particularly important for the detection of GW
bursts from these sources: the energy spectrum of the events and
the distribution of waiting times.

{\em Energy spectrum}. Phenomena related to self-organized
criticality lack an intrinsic energy scale, and for this reason the
number of events $N(E)$ which release an energy  
$E$  has a
power-law distribution
\be\label{GRg}
dN\sim E^{-\g} dE\, .
\ee
This has been verified experimentally for earthquakes, where it is
known as the Gutenberg-Richter law. Of course any energy reservoir,
like the Earth crust, can store only a finite amount of energy, so
\eq{GRg} holds only up to a maximum cutoff energy, above
which $dN/dE$ falls off exponentially. It is remarkable
that the value of the index $\g$ from different
seismically active regions is approximately the same, $\g\simeq 1.6$,
with variation $\pm 0.2$ for different active regions~\cite{GR}.  Even more
striking is the fact that the
distribution of energies of 111 events from the soft gamma repeater
SGR1806$-$20 follows the same
law, with the same value of the index, 
$\g\simeq 1.66$ \cite{Che}. This result has been confirmed
for SGR1627$-$41~\cite{Woods} and 
with much
larger statistics from  observations, by  BATSE and RXTE, of 
 SGR1900+14  \cite{Gog},
which suddenly became extremely active between May 1998
and January 1999, after a long
period of sporadic activity \cite{Hur}. 
With this large statistics (about $10^3$
events) one finds $\g
=1.66\pm 0.05$ over four orders of magnitudes in energy.
A power-law
with the same value of
$\g$ has also been obtained in computer simulations of fractures in a
stressed elastic medium \cite{Katz}.

In GW detectors the crucial problem is to
distinguish a candidate GW burst from a myriad of  spurious
disturbances. To drastically reduce the background
the output of  two detectors are
put in coincidence. The number of
accidental coincidences is estimated 
shifting the data stream of one detector with respect to the other
and counting the number of coincidences (which now
are all accidental). The procedure is repeated for different values of
the time shift, and one finally computes the average over, say,
1000 different shifts~\cite{ROG01}. The result is that, in present
detectors, there is still a significant number of accidentals.
Triple coincidences could give a definitive answer but
require three  sensitive detectors, 
with a good common time of
operation, and this is challenging.
Then, it is clear that
any  further handle that
could help us to discriminate between
a true GW signal and  spurious events is welcome.
For GW bursters, such a handle could be provided by the comparison of
the energy spectrum of the events with
the energy spectrum of  accidental coincidences, which can be measured
experimentally using the shifting algorithm mentioned above. For
instance, thermal noises will follow
a Boltzmann
distribution with some effective noise temperature $T_{n}$, and
therefore at large energies 
$dN/dE\sim\exp\{-E/k_BT_n\}$.

\begin{figure}
\includegraphics[width=0.4\textwidth, angle=0]{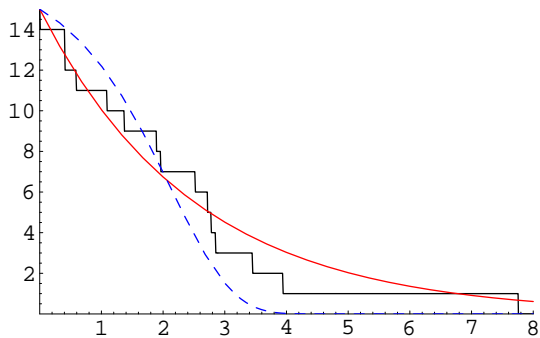}
\caption{\label{fig1} $N(>E)$ as a function of $\log E/E_{\rm min}$.
The black stepwise  line is the function $N(>E)$ generated by
the simulation.
The continuous (red) line  is \eq{pow}
with $k=0.66$ (hence $\g =1.66$), while the dotted (blue)
line is a fit to \eq{expo}. Both curves are normalized so that 
$N(>E)=15$ at $E=E_{\rm min}$.
}
\end{figure}

When looking for a power-law distribution of the events
the main problem will  be the poor statistics
since, even in the most optimistic case, we will have
a small number of candidate GW events. In particular,
it will
be impossible to compare the experimental distribution
with \eq{GRg} performing  a binning in
energy since all bins would be
undersampled. This problem however can be  alleviated considering the
number of events with energy  larger than $E$,
\be\label{N>E}
N(>E)=\int_E^{\infty} dE' \, \frac{dN}{dE'}\, .
\ee
If $dN/dE\sim E^{-\g}$, then
\be\label{Nk}\label{pow}
N(>E)\sim E^{-k}\, ,\hspace{5mm} k=\g -1\, .
\ee
Instead, if $dN/dE\sim \exp\{-E/E_0\}$, then also 
\be \label{expo}
N(>E)\sim \exp\{-E/E_0\}\, , 
\ee
apart from prefactors which will be well beyond
the expected accuracy.
The experimental curve $N(>E)$ changes in steps, decreasing by one
unit each time we reach the energy corresponding to one event.
For illustration, in Fig.~\ref{fig1} we show the result of a
simulation in which we generated 15 events with
 $E\in [E_{\rm min},\infty [$
(where $E_{\rm min}$  defines our units of
energy, and  is the detector threshold)
distributed according to 
\eq{GRg} with $\g =1.66$. In Fig.~\ref{fig1} 
we plot $N(>E)$ as a function of $\log E/E_{\rm min}$. 
The stepwise line   is the function $N(>E)$ generated by
the simulation.
The continuous  line  is \eq{pow}
with $\g =1.66$, while the dotted 
line is a fit to \eq{expo}. This suggests
that, with 15 events, the distinction between a power-like and an
exponential distribution might be possible. 
On real data, of course,
this will have to be quantified with standard statistical tests.
Observe in particular that the exponential curve is
completely unable to account for the existence of one event with
$\log E/E_{\rm min}>7$. 
To distinguish the exponential from the power-law, the crucial role is
of course  played by the most energetic events.

The spectrum of accidentals must be measured experimentally and, 
if it  drops exponentially, 
a strategy to search for
GW bursters is to perform the data analysis with a very high cut on the
energy of the events, since real signals drop only as a power-law. 
If the detectors are so clean
that with these cuts the number of total expected accidentals,
computed from the shifting algorithm mentioned above,  becomes
much smaller than one, then
even the presence of a few coincidences could be significant.

When we look for high-energy events we can relax
other cuts. For instance, in the analysis of data from resonant bars,
presently are taken into consideration only data stretches where the
average effective temperature of the bar is smaller than a given
value, say 6~mK. This is very
reasonable when, as in Ref.~\cite{ROG01}, 
we look for  events that depose in the bar 
more than O(30)~mK (which corresponds to a dimensionless
GW amplitude $h=O(10^{-18})$ 
for a burst with a duration $\sim 1$~ms). 
However, if  we 
restrict to events of
hundreds of mK, the fact that the detector temperature was 6 or 10~mK has
little importance, and we can relax this condition. This has the
consequence that the
stretches of useful data become significantly longer. Furthermore, 
the orientation of the detector with respect to a source
changes with sidereal time because of
the Earth rotation. Low-energy events, which are close to the
detection threshold, can therefore be easily missed when the detector
is not favorably oriented, while high-energy events are efficiently
detected independently of sidereal time.

{\em Waiting time distribution}. The statistics of waiting times
(the times between an event and the next)  of
both earthquakes and SGRs is very different from that of uncorrelated
events. Earthquakes, SGRs and other phenomena related to
self-organized criticality have periods of intense bursting activity, during
which the events arrive in bunches or there is a large event followed by
showers of smaller events; these intense periods are then
followed by long, and sometime
extremely long, periods of quiescence. To quantify this property, it
is convenient to introduce the quantity 
$n(<\tau ;N )$, defined as the number of events with waiting time
smaller than $\tau$, when the total number of events detected is $N$.
Like $N(>E)$ defined in \eq{N>E}, 
this is an integrated quantity, in order to circumvent the same
binning problem discussed for the energy spectrum.
Observe that the waiting time between one event and the next depends
strongly on the resolution that we have for detecting the events: with
a very good resolution we will find many small events which otherwise
would go undetected, and correspondingly the waiting times will be
shorter. Therefore, when we compare the waiting time
statistics of different phenomena
like SGR, earthquakes or GW bursts, we must always perform
the comparison at a fixed value of the total number $N$ of events
detected, taking the $N$ most energetic events from each sample. 
We  normalize $n(<\tau ;N)$ to the total number of
events $N$, defining
$w(\tau ;N)=(1/N)\, n(<\tau ;N)$,
so $0\leq w(\tau ;N)\leq 1$. We also normalize $\tau$ to the total observation
time, so  $0\leq \tau\leq 1$.
A remarkable result is that
SGR and earthquakes have the same waiting time
distribution~\cite{Che}. With the very large statistical sample
from  SGR1900+14  mentioned above, 
it has been shown that their waiting time distributions are compatible with
log-normal functions~\cite{Gog}.

For GW detection, unfortunately, we will rather be in the opposite limit 
of very low number of events. Therefore we now investigate whether the waiting
time distribution for SGR1900+14, taking for definiteness only the 9
more energetic events detected in a six months period in 1998, when the
source was very active, can be  distinguished from the
distribution of uncorrelated events.

The waiting time statistics of uncorrelated events can easily  be
computed analytically: the  probability distribution
$p(\tau ;N)d\tau$ 
for having a waiting
time between $\tau$ and $\tau +d\tau$ when we have a total of $N$ events
(normalized so that $\int_0^1d\tau p(\tau ;N )=1$)
 is a binomial distribution
\be\label{pn}
p(\tau ;N)=N (1-\tau )^{N-1}\, .
\ee
We also checked this result numerically generating random arrival
times. Therefore for randomly distributed arrival times
\be\label{wth}
w_{\rm ran}(\tau ;N)=\int_0^{\tau}d\tau 'p(\tau ';N)=
1-(1-\tau )^N\, .
\ee

\begin{figure}
\includegraphics[width=0.4\textwidth, angle=0]{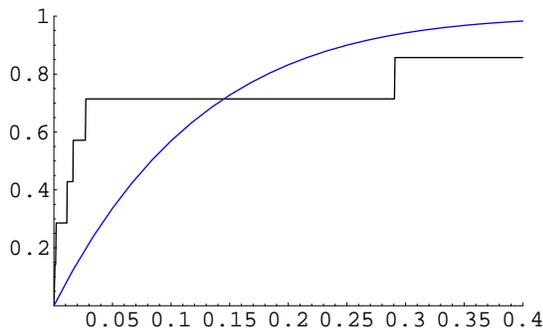}
\caption{\label{fig2} $w(\tau )$ against $\tau$, for the 9 more
 energetic events from
 SGR1900+14 during a 6 months period in 1998 (black stepwise curve)
and $w_{\rm ran}(\tau ;N)$ for $N=9$ (blue continuous line). }
\end{figure}

\noindent
In Fig.~\ref{fig2} we show the experimental distribution for the 9 most
energetic events from  SGR1900+14 \cite{Apt}, 
and we compare it with the function
$w_{\rm ran}(\tau ;N)$ for $N=9$. Even with the uncertainties due the
use of such a limited
sample, we  see that the curve for random events is not compatible
with the experimental data. In particular,
two features  stand
out in the data. First,  at very low $\tau$, the experimental values of
$w(\tau)$ are much {\em higher} that the prediction for random events. For
instance the fraction of events with
waiting times smaller than 
$\tau \simeq 0.03$ (recall that the total observation time
has been normalized to one, so this means waiting times smaller than
$3\%$ of the total observation time)
is over $70\%$ of the total, while
in a random distribution it should be about $20\%$. Physically,
this
reflects the existence of periods of very intense activity, when the
events arrive in bunches. Second, the experimental curve crosses 
$w_{\rm ran}(\tau ;N)$ and then reaches the value $w=1$ at $\tau =1$
(fixed by  normalization) staying {\em below}
$w_{\rm ran}(\tau ;N)$.
Physically, this
reflects the existence of  long periods of quiescence.

This suggest that, even with a rather limited sample,
the  waiting time distribution expected for GW bursters is so peculiar
that it can be
 distinguished from that due to random events.
A different question is whether accidental coincidences in GW
detectors follow the waiting time
distribution of random events. In general, a
certain amount of clustering must be expected because, for
various reasons, in certain periods the detectors can be more noisy.
However, this is an issue that can be studied experimentally. The
distribution of accidental coincidences in two detectors, obtained
from  the shifting algorithm,  can be fully
characterized, and its waiting time statistics can be compared to that
expected for uncorrelated events and for SGRs.

As discussed at length in Ref.~\cite{CDM}, for a source emitting repeatedly
GW bursts another useful tool is a sidereal
time analysis, since the events will be detected with higher efficiency
when the detector is favorably oriented with respect to this source.
This effect will be more important for low-energy events, i.e.
for events close to the detector threshold, while the number of
detected high-energy events
is largely insensitive to the detector orientation. The sidereal time analysis
is therefore complementary to the study of the
waiting time and energy distributions
discussed in this paper.

Finally, all mechanisms that generate GW
bursters are likely to produce x- or $\g$-rays and therefore 
it might be possible a simultaneous detection
of photons and GWs.

{\em Acknowledgments}. We are grateful to Eugenio Coccia for many
useful discussions. Our work is partially supported by the Fond
National Suisse.


\begin{thebibliography}{99}


\bibitem{ROG01} P.~Astone 
{\it et al.},
Class.\ Quant.\ Grav.\  {\bf 19} (2002) 5449.

\bibitem{LIGO}  B. Abbott {\em et al.}, Phys. Rev. {\bf D69}
 (2004) 102001

\bibitem{CDM} 
E.~Coccia, F.~Dubath and M.~Maggiore,
Phys. Rev. {\bf D70} (2004) 084010, arXiv:gr-qc/0405047.


\bibitem{ST} S.L. Shapiro and S.A. Teukolsky, ``Black Holes, White
  Dwarfs and Neutron Stars'', Wiley 1983. p.~12.


\bibitem{DT}
R.~C.~Duncan and C.~Thompson,
Astrophys.\ J.\  {\bf 392} (1992) L9.

\bibitem{TD1} C.~Thompson and R.~C.~Duncan,
Astrophys.\ J.\  {\bf 408} (1993) 194;  {\em ibid.}
{\bf 473} (1996) 322;
Mon. Not. R. Astron. Soc. {\bf 275} (1995) 255.

\bibitem{FP2} J.~A.~de Freitas Pacheco,
Astron. Astrophys. {\bf 336} (1998) 397.

\bibitem{Ioka} K. Ioka, Mon. Not. R. Astron. Soc. {\bf 327} (2001) 639.

\bibitem{Rud} M. Ruderman, Nature {\bf 223} (1969) 597.

\bibitem{SW} R. Smolukowski and D. Welch, Phys. Rev. Lett. {\bf 24} 
(1970) 1191.


\bibitem{DPB} A.~Drago, G.~Pagliara and Z.~Berezhiani,
arXiv:gr-qc/0405145.

\bibitem{MVP} 
G.~F.~Marranghello, C.~A.~Z.~Vasconcellos and J.~A.~de Freitas Pacheco,
Phys.\ Rev.\ D {\bf 66} (2002) 064027.


\bibitem{Bak} P.~Bak, 
{\em How Nature Works}, Oxford Univ. Press,  1997.

\bibitem{GR} B. Gutenberg and C.~F.~Richter, Bull. Seis. Soc. Am. {\bf 46}
  (1956) 105.

\bibitem{Che} B.~Cheng et al.
  Nature {\bf 382} (1995) 518.


\bibitem{Woods} P.~Woods {\em et al.},
Astrophys.\ J.\  {\bf 519} (1999) L139.

\bibitem{Gog} E. G\"og\"us {\em et al.}, 
Astrophys.\ J.\  {\bf 532} (2000) L121.

\bibitem{Hur} K. Hurley {\em et al.}, Nature {\bf 397} (1999) 41.

\bibitem{Katz} J.~I.~Katz, J. Geophys. Res. {\bf 91} (1986) 10412.

\bibitem{Apt} 
R.~L.~Aptekar {\it et al.},
arXiv:astro-ph/0004402.


\end{thebibliography}
\end{document}